\documentclass[lettersize,journal]{IEEEtran}
\usepackage{amsmath,amsfonts}
\usepackage{algorithmic}
\usepackage{algorithm}
\usepackage{array}
\usepackage[caption=false,font=normalsize,labelfont=sf,textfont=sf]{subfig}
\usepackage{textcomp}
\usepackage{stfloats}
\usepackage{url}
\usepackage{verbatim}
\usepackage{graphicx}
\usepackage{cite}
\usepackage{color}
\usepackage{booktabs}
\begin{document}

\title{Semantic Communications Based on Adaptive Generative Models and Information Bottleneck\vspace{.2cm}}

\author{Sergio Barbarossa,~\IEEEmembership{Fellow,~IEEE}, Danilo Comminiello,~\IEEEmembership{Senior,~IEEE}, Eleonora Grassucci,~\IEEEmembership{Graduate,~IEEE}\\\smallskip Francesco Pezone, ~\IEEEmembership{Student,~IEEE,} Stefania Sardellitti,~\IEEEmembership{Member,~IEEE,} Paolo Di Lorenzo,~\IEEEmembership{Senior,~IEEE}
\thanks{Pezone is with Department of Computer, Control and Management Engineering, Sapienza University of Rome, Italy. Barbarossa, Comminiello, Grassucci, Sardellitti, and Di Lorenzo are with the Dept. of Information Engineering, Electronics, and Telecommunications, Sapienza University of Rome, Italy. Email: \{sergio.barbarossa, danilo.comminiello, eleonora.grassucci, stefania.sardellitti, paolo.dilorenzo\}@uniroma1.it. This work was partially supported by the European Union under the Italian National Recovery and Resilience Plan (NRRP) of NextGenerationEU, partnership on “Telecommunications of the Future” (PE00000001 - program RESTART; the PRIN 2017 Project Liquid Edge; the EU-Taiwan project 5G-CONNI, under Grant 861459; and by Huawei Technology France SASU, under agreement N. TC20220919044.} \vspace{-.2cm}}



\maketitle

\begin{abstract}
Semantic communications represent a significant breakthrough with respect to the current communication paradigm, as they focus on recovering the meaning behind the transmitted sequence of symbols, rather than the symbols themselves. In semantic communications, the scope of the destination is not to recover a list of symbols {\it symbolically identical} to the transmitted ones, but rather to recover a message that is {\it semantically equivalent} to the semantic message emitted by the source. This paradigm shift introduces many degrees of freedom to the encoding and decoding rules that can be exploited to make the design of communication systems much more efficient. 
In this paper, we present an approach to semantic communication building on three fundamental ideas: 1) represent data over a topological space as a formal way to capture semantics, as expressed through relations; 2) use the {\it information bottleneck} principle as a way to identify relevant information and adapt the information bottleneck online, as a function of the wireless channel state, in order to strike an optimal trade-off between transmit power, reconstruction accuracy and delay; 3) exploit {\it probabilistic generative models} as a general tool to adapt the transmission rate to the wireless channel state and make possible the regeneration of the transmitted images or run classification tasks at the receiver side.\end{abstract}


\vspace{-.3cm}
\section{Introduction}
\IEEEPARstart{N}{othwistanding} 
the enormous potentials of 5G networks and their expected evolution, significant efforts are already being devoted to designing a future generation of networks, namely Sixth Generation (6G), which is expected to create a cyber-physical continuum between the connected physical world of senses, actions, and experiences and its programmable digital representation. Next-generation networks will need to provide intelligence, limitless connectivity, and full synchronization of the physical and digital worlds. One of the distinguishing features of 6G networks will be the synergy between Artificial Intelligence (AI) and networking: On the one hand, the new networks will need AI to become more autonomous and possibly {\it zero-touch}; on the other hand, the new networks will bring AI services as close as possible to the end users, to satisfy service delay constraints or minimize energy consumption \cite{strinati20216g}.
The new networks will need to be {\it AI-native}, meaning that AI tools will constitute the building blocks of the network itself, rather than just applications running on top of the network. 

One of the main motivations underlying the need for a new generation of communication networks is sustainability: The current exponential growth of data traffic is not sustainable in the long run. To promote sustainability, it is necessary to envisage new communication paradigms able to support new services, but without necessarily increasing the data rates. A possible way out comes from a better understanding of human intelligence and social interaction mechanisms. 


A distinguishing feature of humans with respect to other animals is their apparatus of sophisticated social interaction skills. To communicate with each other and share meaningful experiences, humans create symbols to represent their perceptual experiences. Natural language is the paramount example of human interaction based on the association of sounds to objects of the real world, as well as to abstractions of the real world created by the human brain. Building on the composition of basic sounds, humans form structures, from phonemes to words to sentences, etc., that enable a very complex interaction. A key point to highlight in this association is the trade-off between the complexity of the representation and the relevance of the used symbols to convey the intended meaning, within a certain approximation or distortion. The other fundamental interaction mechanism involves vision. Also in this case, the scene that our brain builds from the signals coming from the optical nerve is a representation of the outside world that is simple enough to convey the relevant information, while discarding a lot of information as irrelevant with respect to the context. 

Both examples show that the association between mental states and symbols (either sounds or visual signs) used to represent them satisfies a fundamental parsimony principle: The association must be simple enough to require small effort and time, while at the same time keeping the relevant information about the intended meaning or the goal of communication.
We believe that next-generation networks should take inspiration from these principles, by focusing on the extraction of the relevant information and the retrieval of the semantics associated with the transmitted symbols. In this paper, we propose an approach to semantic communication over wireless channels based on three important pillars: 
{\it
\begin{itemize}
    \item topology-based knowledge representation,
    \item information bottleneck principle,
    \item probabilistic generative models.
\end{itemize}
}
The first pillar is based on the idea that semantics may be formally represented through the set of {\it relations} between the basic elements of a language, to be intended in a very general way. The formal way to represent relations is to define the associated topological space. Graphs are an important example of a topological space, widely used for example in Natural Language Processing (NLP). However, in general, graphs are unable to capture all the information and it is necessary to resort to higher-order structures, like simplicial or cell complexes, which we assume as the basis for our knowledge representation.

Secondly, we start from the assumption that a communication typically occurs to fulfill a shared goal between source and destination. Then, as the second pillar of our approach, we introduce a variation of the information bottleneck (IB) principle to  identify the information {\it relevant} to the intended goal of communication and we derive the trade-off between complexity and accuracy of the representation. The IB principle is used {\it to guide the design of the source encoder to the goal of communication}, which is not necessarily the recovery of the transmitted bits but rather the fulfillment of a goal. Furthermore, we propose an {\it online mechanism to adapt the bottleneck to cope with the wireless channel variability} in order to induce some desired behavior, like minimization of average energy consumption or average delay necessary to reach a decision on the transmitted data.

However, finding the source encoder from the information bottleneck principle in a general scenario is a rather complex task. For this reason, we propose, as the third pillar, a numerically efficient {\it coding method based on probabilistic deep generative models}. These models are used to represent the relevant parameters to be used at the receiver to regenerate a message that is {\it semantically equivalent} to the transmitted message, using only a reduced subset of symbols. Then, using the IB idea, we adapt the amount of symbols to be transmitted to the channel conditions and to the accuracy in the fulfillment of the communication goal.


The rest of the paper is organized as follows. Section \ref{State of the art} reviews the state of the art. Section \ref{Semantic communication architectures} illustrates a general scheme of semantic communication. In Section \ref{Semantic communications based on topological knowledge representation}, we formulate a formal way to represent semantics through topological spaces, like graphs, simplicial or cell complexes, and show how different spaces can lead to different trade-offs between representation complexity and accuracy.
In Section \ref{Source encoding based on the Information Bottleneck Principle}, we propose a principled way to design a wireless communication system based on a generalization of the  Information Bottleneck principle. Then, in Section \ref{Semantic communication based on Generative Models}, we illustrate a practical application of the IB approach exploiting generative models based on variational autoencoders. 
\section{State of the art}
\label{State of the art}
Current communication systems are still based on Shannon paradigm, which is firmly rooted in probability theory. As already suggested by Shannon's collaborator, Warren Weaver, communication takes place on three levels \cite{weaver1953recent}: 
\begin{enumerate}
    \item {\it ``transmission of symbols (the technical problem)''};
    \item {\it ``semantic exchange of transmitted symbols (the semantic problem)''};
    \item {\it ``effect of semantic information exchange (the effectiveness problem)''}.
\end{enumerate}
Shannon deliberately focused on the technical problem only and derived the fundamental limits for the reliable recovery of the sequence of symbols transmitted by the source after propagation through a noisy channel. Shannon provided also the fundamental rate-distortion tradeoffs, relating the complexity of the representation and the distortion of the reconstruction. Shannon's theory has nothing to do with the meaning that the symbols represent, but only with the frequency of the symbols occurrence. Pressing keys of a computer keyboard at random produces a sequence of symbols that has maximum entropy, i.e., the maximum amount of average information, in the Shannon sense, but it has almost zero semantic information because there is essentially no meaning associated with it. More recently, a Shannon-like rate–distortion theory has been indicated as a principled mathematical framework for understanding human perception and visual perceptual memory \cite{sims2016rate}. But clearly, the formulation of a semantic information theory, based on firm principles, is still an open problem. 
A first step in the search for meaning beyond symbols entails incorporating the logic of the language used to communicate. A first attempt to move in this direction  was developed by Carnap and Bar-Hillel in \cite{carnap1952outline}. Since then, several schools of thought have proposed different alternative approaches to generalize Shannon's information theory, each aimed at emphasizing different perspectives: philosophy of information,  
logic and information,  
information algebra, 
information flow, 
and algorithmic information theory. 

Recently, semantic communication has been identified as a possible breakthrough of 6G networks in \cite{ kountouris2021semantics}. Joint source-channel encoders exploiting the structure of images or natural language have been proposed in \cite{bourtsoulatze2019deep,jankowski2020wireless,xie2021deep}.
In applications such as text transmission, the semantics underlying the text has also been explicitly exploited in designing a joint source/channel coding (JSCC) technique \cite{xie2021deep}. The authors of \cite{weng2021semantic} designed speech recognition-oriented semantic communications to directly recognize the speech signals in texts. More recently, a transformer-based approach has also been investigated in \cite{xie_letaief2021task} to support both image and text transmission. Alternative methods were also proposed in \cite{farshbafan2021common} to define an optimized \emph{common-language} between a listener and a speaker, employing reinforcement learning and curriculum learning. Very recently, a semantic framework for video conferencing has been proposed, by exploiting semantically relevant keypoints representing the facial expressions \cite{jiang2022wireless}.

\section{Semantic communication architectures}
\label{Semantic communication architectures}
\begin{figure*}
\centering\includegraphics[width=0.95\textwidth]{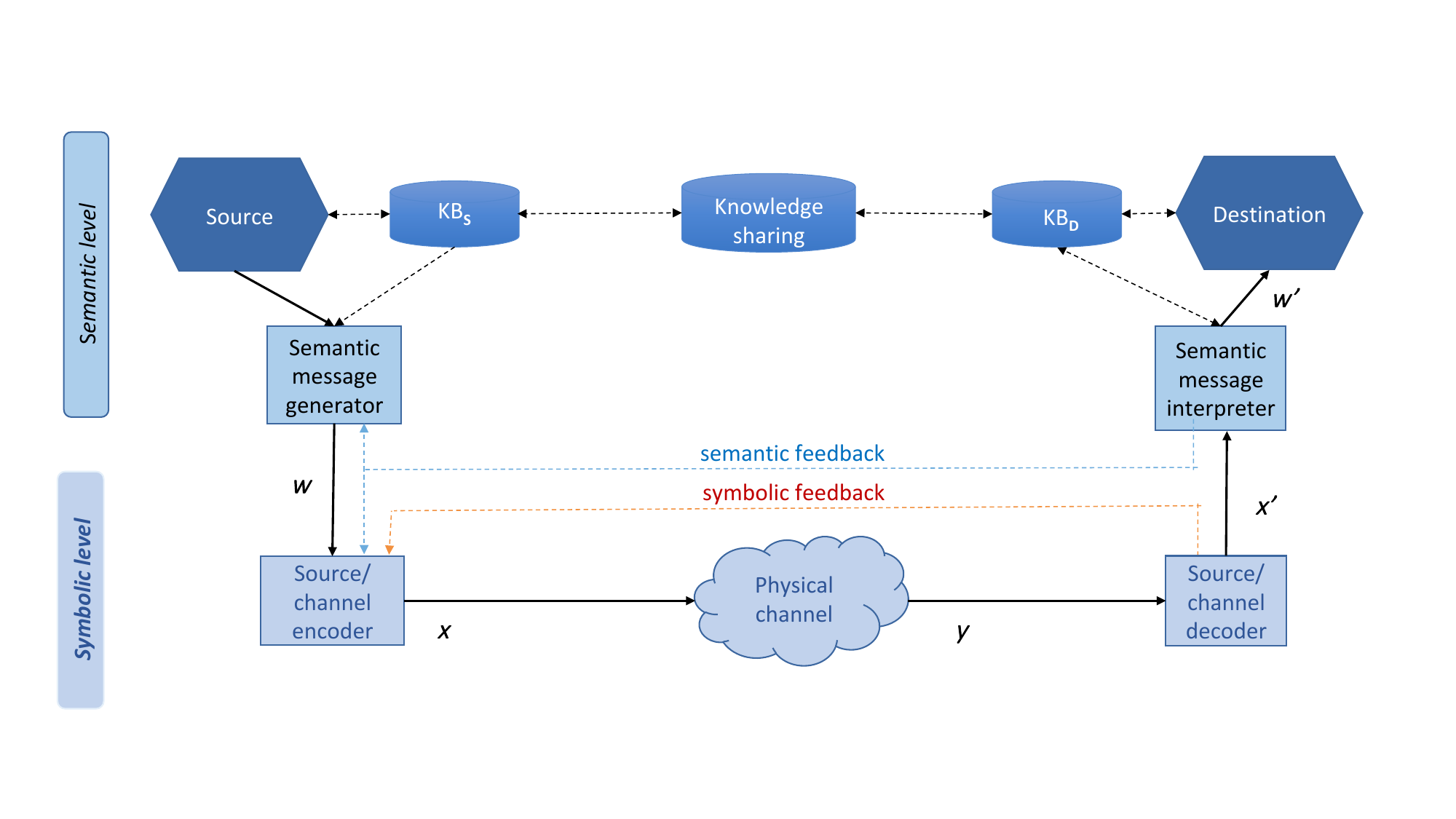}
\caption{Block diagram of a semantic communication system.}
\label{fig:block diagram}
\end{figure*}
A possible semantic communication system is sketched in Fig. \ref{fig:block diagram}, where the semantic level is represented on top of the symbolic level. In this block diagram, it is important to distinguish the {\it semantic message} ${\bf w}$ from its symbolic representation ${\bf x}$, which we denote as the {\it symbolic message}. 
The distinguishing feature of semantic communication, as opposed to conventional (i.e., symbolic) communication,  is that, whichever is the set of symbols used by the source to represent and communicate a semantic message, the scope of the receiver is not to recover the sequence of transmitted symbols, but only to recreate a message that is {\it semantically equivalent} to the transmitted one. This means that there is no (semantic) error even if the symbolic message $\mathbf{x'}$ regenerated at the receiver side is completely different from $\mathbf{x}$, as far as the corresponding symbolic message $\mathbf{w'}$ is semantically equivalent to $\mathbf{w}$. This property is what makes semantic communication potentially much more efficient than conventional communication because it adds many degrees of freedom in the reconstruction process. This advantage is paid in terms of the complexity needed to represent semantics.

As depicted in Fig. \ref{fig:block diagram}, the mapping from a semantic message to its symbolic representation occurs via a knowledge base (KB) system, which incorporates the rules underlying the language used to communicate \footnote{Ideally, source and destination should possess the same KB. In practice, the two KB systems can be different, as in Fig. \ref{fig:block diagram}, but a knowledge-sharing mechanism needs to be implemented to  adapt the destination KB to the source KB.}.
A language, in very general terms, is composed of a set of elements (objects) and a grammar, i.e. a set of rules used to combine the elements in order to create messages that have a meaning in the given language. A KB system is typically composed of a computational ontology, facts, rules, and constraints \cite{chein2008graph}.  
As we will see in Section \ref{Semantic communication based on Generative Models}, a possible implementation of a knowledge base system may be given by a variational autoencoder (VAE). 

An important feature of the scheme of Fig. \ref{fig:block diagram} is the presence of a feedback from the destination to the source. Feedback can occur {\it within} levels (i.e. technical or semantic level), or {\it across} levels, i.e. from the semantic to the technical level. This feedback is important because it makes possible to adapt the transmission rate as a function of how well the destination is able to take correct decisions about the semantics conveyed by the symbols or to fulfill the goal of communication.
In Section \ref{Source encoding based on the Information Bottleneck Principle}, we will show how to implement this feedback by adapting the IB to the channel variability and to the accuracy of the decision at the destination side. 
\section{Knowledge representation over topological spaces}
\label{Semantic communications based on topological knowledge representation}
\begin{figure}[t]
\centering
\hspace{1cm}\includegraphics[height=5.3 cm,width=7.9cm]{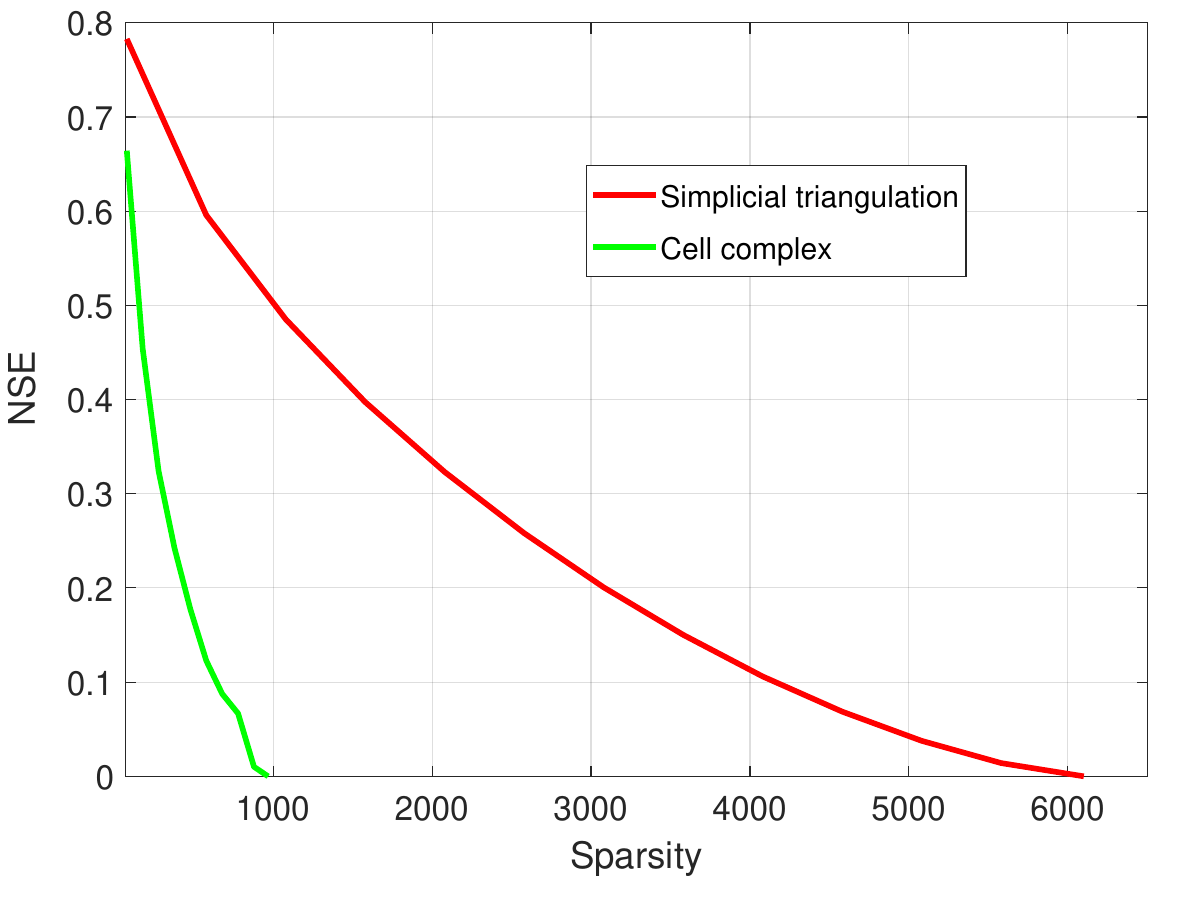}
\vspace{.2cm}
\caption{Accuracy vs. representation complexity.}
\label{fig:Graph_semantic_encoding}
\end{figure}

A shared definition of semantics is still missing. Nevertheless, we may think of representing semantics through the set of {\it relations} between the elements of a language. The formal way to represent relations is by introducing the corresponding {\it topological space}, i.e. a space composed of a set of elements along with a set of relations among them. The simplest example of a topological space is a graph, representing {\it pairwise or dyadic} relations (edges) between pairs of objects (vertices). Graph-based knowledge representation mechanisms have been largely investigated for example, in Natural Language Processing (NLP), where graphs play a key role in associating a meaning to a sentence and in solving disambiguation problems.  
In computer vision, semantic relationships are typically expressed as graphs grouping portions of the image representative of the same object class, or to recognize the shape of 3D objects. However, graphs are limited as they are only able to encode  {\it pairwise} relations.
In many applications, the semantic relations may not be restricted to pairwise relations, but should incorporate higher-order (multiway) relations. Examples 
of higher-order graphs, still endowed with a rich algebraic structure, are simplicial or cell complexes. In most applications, the data to be transmitted are defined over a topological space. 
The idea we wish to highlight is that the identification of the space, through the relations among its constitutive elements, 
has a deep impact on our capability to find a data representation that yields a good trade-off between accuracy and complexity.
To investigate how  higher-order topological spaces can successfully encode the semantic  relationships among the observed data, we consider as a numerical example an image segmentation task. The goal is to find a semantic segmentation of the image based on cell complexes by associating each segment to a polygonal cell over which the same semantic information  is observed.  We considered a gray-scale version of Mona Lisa portrait of Leonardo da Vinci, represented
using $12$ gray levels. As a first example of space, we build a gradient-based Delaunay
triangulation of the image using $6100$ triangles. The signal associated with this space is the image intensity over each triangle. Then, as a second example of space, we glued all triangles having similar image intensity, thus forming a cell (polygonal)
complex. In the example, the resulting cell complex is composed of $969$ cells.
Then, to compare the representations over different spaces, we built a dictionary using the eigenvectors of the Laplacian matrices associated with each space and computed the optimal trade-off between the accuracy of the signal representation over each dictionary and the corresponding complexity (number of dictionary elements used to represent the image). In Fig. \ref{fig:Graph_semantic_encoding},  we  plot the accuracy (normalized squared error (NSE)) vs. complexity. From Fig. \ref{fig:Graph_semantic_encoding}, it is evident the considerable gain in terms of accuracy/complexity trade-off achievable using a cell complex instead of a simplicial complex.

\section{Source encoding based on the Information Bottleneck Principle}
\label{Source encoding based on the Information Bottleneck Principle}
In this section, we dig further into knowledge representation and put it in the context of goal-oriented communication over a wireless channel. An example can be an IoT device that sends images to a server that runs an image recognition application. Denoting by $X$ the transmitted vector (image)
and by $Y$ the corresponding decision (label) to be taken at the receiver side, our first task is to identify the information strictly {\it relevant} to the fulfillment of the (decision) goal. Our second task is to derive the structure of the source encoder that extracts this relevant information while achieving the best trade-off between complexity of the representation (number of bits used to represent the relevant information) and accuracy on the decision variable $Y$. 
To approach this problem formally, we introduce a variation of the Information Bottleneck (IB) principle \cite{tishby2000information} as a basic principle to design the source encoder in order to minimize the complexity (number of bits) of the representation, while preserving only the {\it relevant} information about $Y$. More specifically, denoting by $T$ the compressed variables, the conventional IB method finds the probabilistic compression rule, given by the conditional probability $p_{T/X}(t/x)$, that minimizes a linear combination, with nonnegative coefficients, of the mutual information $I(X; T)$ between the input $X$ and the encoded variable $T$ and the mutual information $I(T; Y)$ between and the encoded variable $T$ and the decision variable $Y$. In this formulation, $I(X; T)$ represents the complexity of the encoding rule (measured in number of bits),  while $I(T; Y)$ represents the {\it relevance} of the encoded variable $T$ for the retrieval of $Y$. In this paper, we modify the IB principle to make it suitable for transmission over wireless communications. First of all, we consider the real case where the transmitted data are corrupted by noise, so that the receiver does not have direct access to $T$, but only to a version $T + \eta$ corrupted by the channel noise $\eta$. Second, we redefine the bottleneck optimization problem as 
\begin{equation}
    \min_{A, M} \;\; I(X; T) + \beta \cdot MSE(Y, \hat{Y}),
    \label{eq:IB main problem}
\end{equation}
where $T = A \cdot X + \xi$ denotes a linear encoder, including additive noise $\xi$, while $\hat{Y} = M \cdot (T + \eta)$ is a linear estimate of $Y$; $\beta$ is a scalar parameter that allows us to tune the trade-off between complexity and relevant information: small values of $\beta$ lead to small complexity encoders, but possibly large distortion, whereas large values of $\beta$ yield a small distortion, but with greater complexity. Even though the new optimization problem in \eqref{eq:IB main problem} is nonconvex, we managed to find a closed-form solution for the case where all the involved variables are Gaussian. The linear structure of the encoder/decoder is optimal in the Gaussian case, but suboptimal in general. For this reason, in Section \ref{Semantic communication based on Generative Models} we will use a more general structure based on variational autoencoders.\\

\begin{figure}[t]
    \centering
\includegraphics[width=\linewidth]{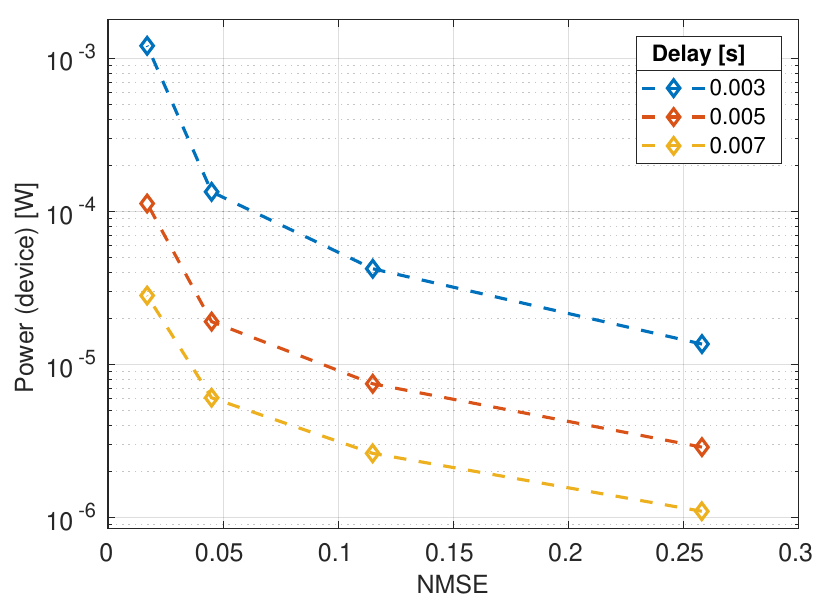}
    \caption{Trade-off between Power-Delay-Error}
  \label{fig: trade-off lyapunov}
\end{figure}

Our idea is to adapt $\beta$ as a function of the channel state and of the required classification task accuracy, while also enforcing a constraint on the decision delay, defined as the sum of the processing and communication delay.
To cope with the unpredictable wireless channel variability, we adopt a stochastic optimization framework, able to work without knowledge of the channel probabilistic model and of the task arrival rates. In this framework, time is slotted in intervals of equal duration. In each time slot, the problem is converted into the optimization of a deterministic problem, for each time slot, taking into account what has been done in the previous slot by introducing a set of real and virtual queues that quantify how well the method is satisfying the optimization constraints. The optimization variables are the transmit power of all the devices, the clock rates used at both source and destination sides to process the data, and the trade-off parameter $\beta$ used to adjust the bottleneck. The objective of the optimization is to minimize the average energy consumption, under constraints on average accuracy and service delay. A numerical result is reported in Fig. \ref
{fig: trade-off lyapunov}, showing the average power consumption vs. the Normalized Mean Square Error (NMSE), in a regression problem, for three different values of the average delay. The delay is composed of the sum of the transmission delay plus the time necessary to process the data. As expected, imposing smaller average delays forces the system to use more power to achieve the same NMSE value. 
The above results refer to a simple regression task and adopt linear encoders because they represent the optimal choice in the Gaussian case. In the next section, we generalize the idea to the general case, implementing the IB principle using a variational autoencoder. 

\section{Semantic communication based on\\ Generative Models}
\label{Semantic communication based on Generative Models}
Generative models are experiencing an increasing interest in the AI (and not only) community due to their impressive results in a plethora of tasks including image super-resolution, denoising, image-to-audio translation, 3D synthesis. 
The core idea of generative modeling is to build a sufficiently complex and powerful model for learning the data distribution so as to acquire the ability to draw representative samples from the learned distribution. When learning the distribution, the model grasps the semantic information of the samples and it is able to generate samples with similar semantic information. As an example, when learning to generate human faces, the neural model begins generating semantic aspects of them, such as eyes, nose, and mouth, while progressively building also other contents such as hair and ears.

In this work, we use generative models as the basic building block of a semantic communication system. 
\begin{figure*}
    \centering
    \includegraphics[width=\textwidth]{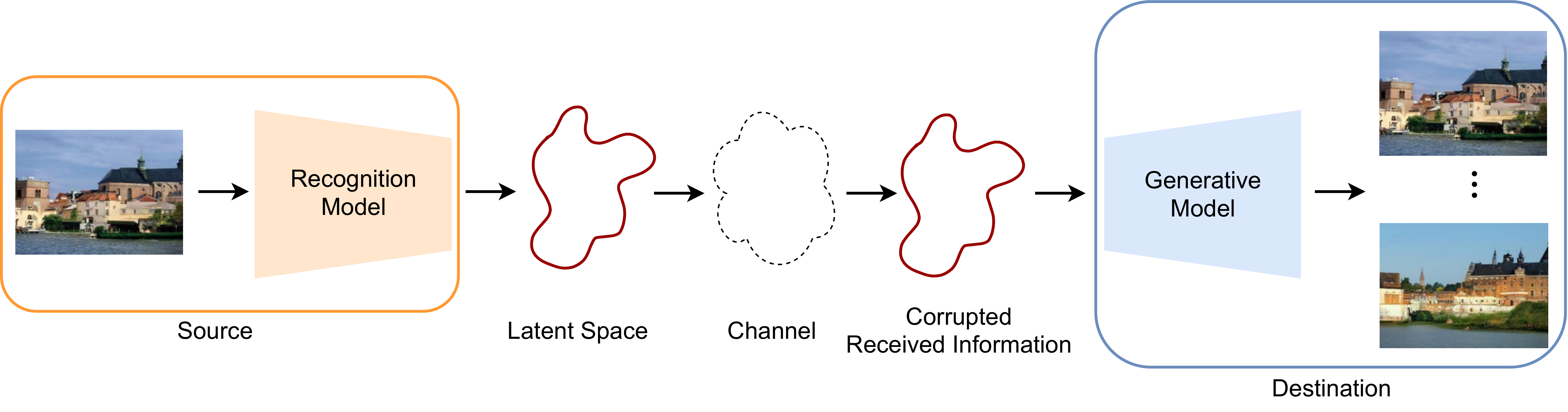}
    \caption{A generative semantic communication framework based on latent variable models.}
    \label{fig:inv_problem_generative}
\end{figure*}
In particular, we focus on a broad class of generative models called \textit{generative latent variable models} (GLVMs) \cite{NEURIPS2021Vahdat}, 
which first transform the input data into a low-dimensional space, the latent space, containing the semantic characteristics of such data, and then sample from this space to generate data having semantically similar features in the data space. GLVMs are composed by a pair of deep neural networks: i) an \textit{encoder}, or \textit{recognition model}, whose objective is to learn the \textit{latent} variables that contain the semantic features of the input, and ii) a deep neural network called \textit{decoder}, or \textit{generative model},  whose aim is to transform the received latent representation into new reconstructed samples with preserved semantic characteristics. In our approach, the two blocks of a GLVM are used as the encoder and decoder components of our communication scheme, as depicted in Figure \ref{fig:inv_problem_generative}.

In our semantic communication scheme the transmitted symbols are the  latent variables, which live in a low-dimension space, while the receiver uses the received variables to regenerate an image using the generative model. The recognition and generative models are trained jointly offline, incorporating a non-trainable channel model in the middle, as to simulate the presence of channel noise during the training phase. The scheme of Fig.  \ref{fig:inv_problem_generative} has enormous potentials in terms of source compression with respect to conventional schemes, and thus reduction of the transmit data rate, because it exploits the shared knowledge between transmitter and receiver. This information is encoded in the GLVM. The advantage comes from the fact that the generated image at the destination side may not coincide in any single pixel with the transmitted image, and yet it may well represent the transmitted image. 
The distinguishing feature of a GLVM used in this way is that {\it the dimension of the latent space can be adapted over time, depending on channel conditions, without retraining the two networks}. The scheme of Fig. \ref{fig:inv_problem_generative} is an example of the general scheme of Fig. \ref{fig:block diagram}, where the source and destination KB's are the recognition and generative models, and the feedback from destination to source is given by a parameter that represents the channel state (symbolic feedback) and a parameter that represents the goodness of the task implemented at the receiver (semantic feedback), like image classification or image regeneration: when the channel is very good, more parameters are sent, thus enabling an accurate recovery; conversely, when the channel is very poor,  fewer parameters are transmitted, but they are still the most relevant ones.
\begin{table}[t]
\centering
\caption{Results for dynamic and static channel conditions under different PSNR values during the transmission.}
\resizebox{\linewidth}{!}{
\begin{tabular}{@{}ccc|cc@{}}
\toprule
&&& \multicolumn{2}{c}{Classification Accuracy$\uparrow$} \\
Channel condition & PSNR (dB) & Latent dim & CIFAR10 & MNIST  \\ \midrule
No channel & - & - & 92.701 &  98.880 \\
Static & 100   & 4, Fixed & 88.262 & 96.433  \\
Static & 25    & 4, Fixed & 87.804 & 97.381  \\
Static & 15    & 4, Fixed & 87.982 & 97.648  \\
Static & 12.5  & 4, Fixed & 88.238 & 97.266  \\
Static & 10    & 4, Fixed & 86.934 & 96.273  \\
Dynamic & Varying & 3, Varying & 86.840 & 96.305  \\ \bottomrule
\end{tabular}}\label{tab:vae_results}
\end{table}

\begin{figure}
    \centering
    \includegraphics[width=\linewidth]{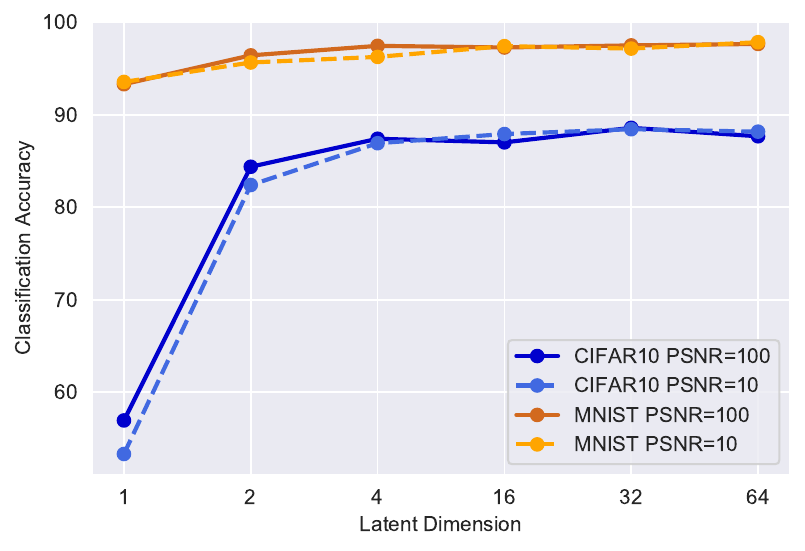}    \caption{Classification accuracy with channel conditions equal to PSNR=100 and PSNR=10 varying the latent dimension from $1$ to $64$.}
    \label{fig:acc_for_latent_dim}
\end{figure}

As an example of application, we evaluated the proposed architecture under different channel conditions and using two image benchmarks, namely MNIST and CIFAR10. 
The goal of the receiver is to classify the images by processing the received latent variables. However, the receiver does not have access to the true latent variables, as they are corrupted from the propagation through the wireless channel. Then we tested how robust is the generative model, and as a consequence the classification task, as a function of the distortion, assessed through the  Peak-Signal-to-Noise-Ratio (PSNR), defined as the ratio between the maximum power of a signal and the power of the corrupting noise. Furthermore, to simulate the variability of the wireless channel, we conducted our tests on a dynamic channel scenario, where the channel varies at random over different time slots. To simulate channel random fading, we vary the PSNR randomly over different time slots. In the static channel condition, we always transmit the same amount of information by fixing the dimension of the latent vector to $4$, for each benchmark. The information is therefore only degraded due to the channel noise, but it maintains the same dimension. On the contrary, under dynamic channel conditions, we dynamically adapt the dimension of the latent vector to be sent as a function of the channel state. Table \ref{tab:vae_results} shows the results of the experimental evaluation as a function of the PSNR, while Figure \ref{fig:acc_for_latent_dim} shows the classification accuracy for different latent dimensions.
From Table \ref{tab:vae_results}, we observe that the classification accuracy decreases as the PSNR decreases, as expected, but the scheme is rather robust against noise. It is worth noting that the dynamic adaptation of the latent space dimension achieves similar performance, even sending fewer latent variables, on average. However, the most striking feature of the proposed architecture is that the system works quite well even sending very few data. This aspect is made more evident in Fig. \ref{fig:acc_for_latent_dim}, where we plot the classification accuracy vs. the dimension of the transmitted latent vector, for two different values of the PSNR.
From the figure it is evident how, in the MNIST case, the number of features to be sent to achieve good classification accuracy can be extremely low. The situation is a bit more complex with the CFAR 10 data-set, but still the number of data to be sent is strikingly low compared to the original images.
Again, we can observe a relative robustness against channel noise.

\section{Conclusion}
In this paper, we have proposed an approach to semantic communications based on three main tools: knowledge representation using topological spaces; design of the source encoder based on the information bottleneck; use of a variational autoencoder to apply the information bottleneck to the most general case. More specifically, we used a generative latent variable model as a way to extract semantic information in the form of latent variables. The distinguishing feature of our scheme is that we can adapt the number of variables to be transmitted, depending on the wireless channel state and on the accuracy of the decisions taken at the receiver side. Furthermore, we tested the robustness of the generative part when working on a set of latent variables distorted by the propagation through a wireless channel. 




\bibliographystyle{IEEEbib}

\begin{IEEEbiographynophoto} {Sergio Barbarossa}  is a Full Professor at Sapienza University of Rome and a Senior Research Fellow with the Sapienza School of Advanced Studies, Rome, Italy. He is an IEEE Fellow and a EURASIP Fellow. His current research interests include semantic and goal-oriented communications, topological signal processing and learning. 
\end{IEEEbiographynophoto}


\begin{IEEEbiographynophoto}{Danilo Comminiello}
is an Associate Professor with the DIET Dept. at Sapienza University of Rome, Italy.
His research interests concern the design of modern artificial intelligence algorithms, including neural networks and generative models.
He is an elected member of the IEEE Machine Learning for Signal Processing Technical Committee.
\end{IEEEbiographynophoto}

\begin{IEEEbiographynophoto}{Eleonora Grassucci}
is currently an assistant professor at DIET Dept. of Sapienza University of Rome, Italy. She was awarded of the Best Track Manuscript Recognition by the IEEE Circuits and Systems Society in 2022.
\end{IEEEbiographynophoto}

\begin{IEEEbiographynophoto}{Francesco Pezone} is a Ph.D. student in Data Science at Sapienza University of Rome, Italy, jointly with the Technical University of Berlin, Germany. His current research interests include Semantic Communication and Deep Learning for Data Compression.
\end{IEEEbiographynophoto}
\begin{IEEEbiographynophoto}{Stefania Sardellitti}  is an assistant professor at Sapienza University of Rome. She received the 2014 and the 2020 IEEE Best Paper Awards from the IEEE Signal Processing Society.  Her current research interests are in the area of  topological signal processing.
 \end{IEEEbiographynophoto}

\begin{IEEEbiographynophoto}{Paolo Di Lorenzo} is an Associate Professor at Sapienza University of Rome. His research interests include topological signal processing, distributed optimization, wireless edge intelligence, goal-oriented and semantic communications. He is recipient of the 2022 EURASIP Early Career Award. He has received three best student conference paper awards sponsored by the IEEE signal processing society and EURASIP.
\end{IEEEbiographynophoto}


\vfill

\end{document}